\pdfoutput=1
\documentclass[
               aps,
               prl,
               twocolumn,
               superscriptaddress,
               nofootinbib,
               showpacs]{revtex4}
\usepackage[utf8]{inputenc}
\usepackage{graphicx}

\usepackage{hyperref}
\usepackage{amsmath}
\usepackage{amssymb}
\usepackage{amsfonts}
\usepackage{latexsym}
\usepackage{pifont}

\usepackage{color}


\begin{document}

\title{Photon Number Statistics of Multimode Parametric Down-Conversion}
\date{\today}

\author{M. Avenhaus}
\email[E-mail: ]{mavenhaus@optik.uni-erlangen.de}
\affiliation{Max-Planck Research Group, G\"unther-Scharowsky-Stra\ss{}e 1/Bau 24, 91058 Erlangen, Germany}
\author{H.~B. Coldenstrodt-Ronge}
\affiliation{Clarendon Laboratory, University of Oxford, Parks Road, Oxford, OX1 3PU, United Kingdom}
\author{K. Laiho}
\affiliation{Max-Planck Research Group, G\"unther-Scharowsky-Stra\ss{}e 1/Bau 24, 91058 Erlangen, Germany}
\author{W. Mauerer}
\affiliation{Max-Planck Research Group, G\"unther-Scharowsky-Stra\ss{}e 1/Bau 24, 91058 Erlangen, Germany}
\author{I.~A. Walmsley}
\affiliation{Clarendon Laboratory, University of Oxford, Parks Road, Oxford, OX1 3PU, United Kingdom}
\author{C. Silberhorn}
\affiliation{Max-Planck Research Group, G\"unther-Scharowsky-Stra\ss{}e 1/Bau 24, 91058 Erlangen, Germany}

\begin{abstract}

We experimentally analyze the complete photon number statistics of parametric down-conversion and ascertain the influence of multimode effects. Our results clearly reveal a difference between single mode theoretical description and the measured distributions. Further investigations assure the applicability of loss-tolerant photon number reconstruction and prove strict photon number correlation between signal and idler modes.

\end{abstract}

\maketitle



The photon statistics of a light beam are the ``fingerprint'' of its quantum
state, from which a number of useful measures of non-classicality may be
inferred. Coherent states reflect closest the classical properties of optical
fields. These states exhibit a Poissonian photon number distribution, while
sub-Poissonian statistics are a clear signature of quantum light. On the other
hand, single-mode thermal radiation shows super-Poissonian photon
statistics. However, thermal light sources typically emit a multimode field so
that their photon statistics are given by the convolution of several thermal
photon number distributions, which, in turn, leads again to a Poissonian
distribution. In some applications, such as quantum key distribution, the
photon statistics and their interplay with the multimode structure are
intrinsically important for determining the security of the communication.

For multi-partite systems another important quantum characteristic is the
extent to which they are correlated beyond what is possible classically.
Quantum correlations, or entanglement, underlie much of the information
processing power of quantum systems. For multiple light beams it is therefore
important to determine the correlations between the different modes, and
photon-detection methods that can reveal these correlations become paramount.
The joint photon statistics of two beams can provide a useful measure of
correlation and the degree of entanglement, also important for applications
such as decoy state quantum cryptography \cite{Mauerer:2007p2297}.

In this Letter we report, to our knowledge, the first direct detection of
complete higher order photon number correlations and statistics between two
light beams that are strictly correlated in photon number (a twin-beam), using
a photon-number-resolved characterization whose performance is compromised
neither by dark-count noise nor by detector inefficiencies. This enables us to
detect strict photon number correlations for all observed photon numbers
including the vacuum. We are also able to identify features caused by the
multimode character of the twin-beam source used. 

In general, photodetectors are not intrinsically mode-selective, since they
have a finite \'etendue and spectral response bandwidth. In this regard they
are ideal for measurements in the photon-number basis. However, detectors that
are sensitive enough to respond to single photons are usually unable to
resolve the differences between small numbers of incident photons. In
contrast, detectors for large photon fluxes usually have high efficiency, but
are too noisy to be sensitive to individual photons. Hence, there currently
exist two main approaches to recover photon number statistics.

The first approach, homodyne detection, employs an ancillary optical beam, the
local oscillator, to increase the photon flux level so that it may be
registered on a standard photodiode. The photon statistics needs to be
recovered indirectly from a large data set by means of quantum state
tomography \cite{Breitenbach:1997p794} or the use of pattern functions
\cite{Leonhardt:1995p3736}. Therefore the photon number statistics are only
recovered indirectly involving the acquisition of a large data set. Further,
the local oscillator necessarily acts as a mode filter, picking out the
component of the input beam that matches the mode of the ancilla. Hence, the
multimode character becomes obscured by the detection technique. Although
multimode homodyne detection is possible \cite{Dorrer:2001p3740}, it is
laborious and requires a very high signal to noise ratio to accurately
estimate the photon numbers.

The second approach combines mode-multiplexing with single-photon sensitivity
to allow for multimode state characterization without the need of a local
oscillator. However, achieving single-photon sensitivity, high quantum
efficiency and low dark count noise constitutes an experimental challenge. The
two most common types of detectors for such applications are avalanche
photodiodes (APDs) and visible-light photons counters (VLPCs). VLPCs can
detect single photons and achieve multi-photon resolution by spatial
mode-multiplexing. They have been used to determine the non-Poissonian photon
number statistics of degenerate parametric down-conversion (PDC), to analyze
the statistics of heralded number states using twin beams, and to verify the
non-classicality of the PDC states \cite{Waks}. While VLPCs exhibit very high
quantum efficiencies, they suffer from significant dark count contributions
and require cryogenic cooling. Similar difficulties are encountered using
sensitive CCD cameras at room temperature \cite{Haderka2005}. Photon number
resolved measurement can also be achieved combining the idea of
mode-multiplexing with the ease of operation offered by commercial APDs. Here
an input pulse is distributed in the time domain into multiple temporal modes
using an optical fiber network \cite{Fitch_Achilles}. In 2006 Achilles
\emph{et al.}~introduced a new technique for loss-tolerant characterization of
photon number statistics \cite{Achilles} using a time multiplexed detector
(TMD) to characterize the conditioned statistics of one- and two-photon states
from a twin beam.

Generation of twin beams by means of PDC has become a widespread way to
implement highly correlated quantum states. Sub- and super-Poissonian photon
number statistics \cite{DeRiedmatten_Rarity}, photon anti-bunching
\cite{Koashi:1993p1390} and quantum correlated quadrature amplitudes
\cite{Rarity:1992p346} have been demonstrated. More recently, a PDC source was
employed to prepare "kitten" states by means of photon subtraction
\cite{Ourjoumtsev:2007p2824}, paving the way for a new approach to conditional
quantum state synthesis.

Although some measurements of joint photon number statistics of PDC have been
performed, a complete and accurate measurement of the joint photon number
probability $p_{n,m}$ for two beams had not been achieved. Such measurements
should quantify the photon number correlations between the signal and idler
beams to all orders, and determine the moments of the unconditional (marginal)
and conditional photon distributions for each beam. Here, we show such a set
of measurements. We employ two TMDs for detecting simultaneously signal and
idler beams, for a range of parametric gains. We are able to build up the full
joint, marginal and conditional photon number distributions, and to confirm
various important correlations. The consistency of the reconstructed
probability distributions across a variety of measurements validates our
approach, and enables us to measure directly the photon number distribution
even for highly multimode beams.

Most crucial for loss-tolerant reconstruction of the photon number
distribution of a twin beam state $|\Psi \rangle = \sum_n \sqrt{p_n} |n,n
\rangle$ is a knowledge of losses, or equivalently the detection efficiency.
Calibration according to the extended Klyshko method \cite{Fitch_Achilles,
Achilles} requires the a priori assumption of photon number correlation. This
assumption, however, needs verification, and we are able to show this
experimentally using the TMDs.

At the heart of our experiment is a TMD, whose response \cite{Achilles:2004p2997}
for incoming photon statistics $\vec{p}=(p_0, \ldots, p_n, \ldots)$ is modeled
as
\begin{eqnarray}
	\label{eq:singleTMD}
	\vec{\varrho} & = & \mathbf{C} \mathbf{L}(\eta) \vec{p} \\
	\mathbf{L}(\eta) = L_{n,m} & = & \dbinom{m}{n} \eta^n (1-\eta)^{m-n}.
\end{eqnarray}
The mathematical description includes a loss matrix $\mathbf{L}(\eta)$ and a
convolution matrix $\mathbf{C}$. The loss relates the individual survival
probability $\eta$ of one photon to the survival of $n$ out of $m\geq n$
photons. The convolution accounts for a probabilistic distribution of $n$
photons into $k$ different time-multiplexed modes whilst taking into account
an individual population probability $P_k$. Both matrices act on a photon
probability vector $\vec{p}$ and yield an observable click vector
$\vec{\varrho}$ in the detector. The probability matrices $\mathbf{L}(\eta)$
and $\mathbf{C}$ act non-deterministically on the measurement outcome on a
shot-by-shot basis. Nevertheless, they provide an analytic and deterministic
functional dependence for state reconstruction when applied to an ensemble
measurement. For accurate photon number measurements, all model parameters
$\eta, P_k$ and $\vec{\varrho}$ need to be experimentally accessible.

\begin{figure}[htbp]
    \centering
\includegraphics[width=\linewidth]{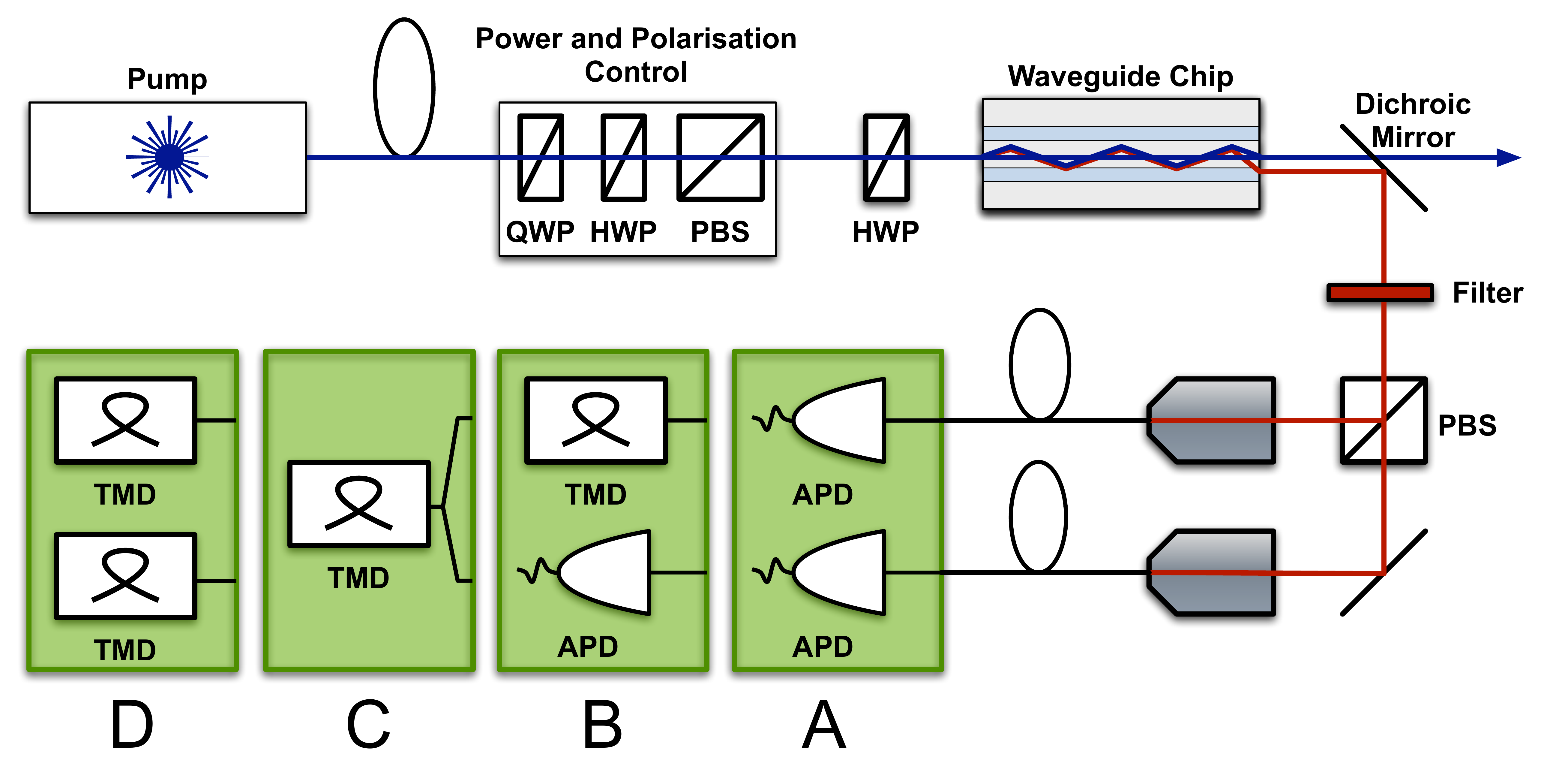}
		    \caption{Scheme of experimental setup (see text);  measurement 
		    configurations A, B, C and D provide information about detector 
		    efficiency, marginal and collective statistics, and joint correlations 
		    respectively.}
    \label{fig:OpticalSetup}
\end{figure}

Our experimental setup is depicted in Figure \ref{fig:OpticalSetup}. A diode
laser source at 404nm pulsed with a repetition rate of 1MHz, with spectral and
temporal bandwidths of 2.0nm and 60ps FWHM respectively, undergoes mode
cleaning, followed by polarization control. It is subsequently coupled into a
periodically poled $\text{KTiOPO}_4$ waveguide chip by means of a focussing
graded index lens. The nonlinear interaction of pump photons decaying into
signal and idler photons via a type-II PDC process (at central wavelengths
$\lambda_s = 740$nm, $\lambda_i = 890$nm) results in the generation of
orthogonally polarized, photon number correlated twin beams.

After state preparation, a dichroic mirror in combination with a red pass
filter splits the PDC state from the strong pump. A polarizing beam splitter
separates signal and idler modes, which are then coupled into multimode fibers
for detection with different APD or TMD measurement configurations, labeled as
A,B,C and D in Figure \ref{fig:OpticalSetup}. These setups are fiber pigtailed
for easy interchangeability with minimal influence on the detection
efficiency. Tight temporal filtering with precision of 82ps is applied during
data acquisition for noise reduction. The experiments are performed at
cw-equivalent powers of pump light coupled through the waveguide, ranging from
40nW to 375nW.

The different measurement configurations are used for a sequence of
experiments that allow for checking the consistency of the recovered
information about the state. We begin with determining the detection
efficiency and then use this calibration to analyze the marginal statistics of
signal and idler mode respectively. The measurement of the collective
statistics of both modes is then followed by the complete measurement of the
joint photon number statistics of the PDC source.


Setup A provides the basis for measuring the efficiency $\eta_K =
\frac{R_C}{R_{S}}$ from the count rates of coincidence ($R_C$) and single
($R_S$) events. We calibrate the real efficiency $\eta$ by evaluating $\eta_K$
at the lowest pump power of 40nW and observe signal and idler efficiencies of
$\eta_s = 11.7\%$ and $\eta_i = 13.7\%$ respectively. Our PDC shows high
source brightness even at this low pump power, yielding count rates of 4950/s
and 5764/s for signal and idler.

\begin{figure}[htbp]
    \centering
		\includegraphics[width=\linewidth]{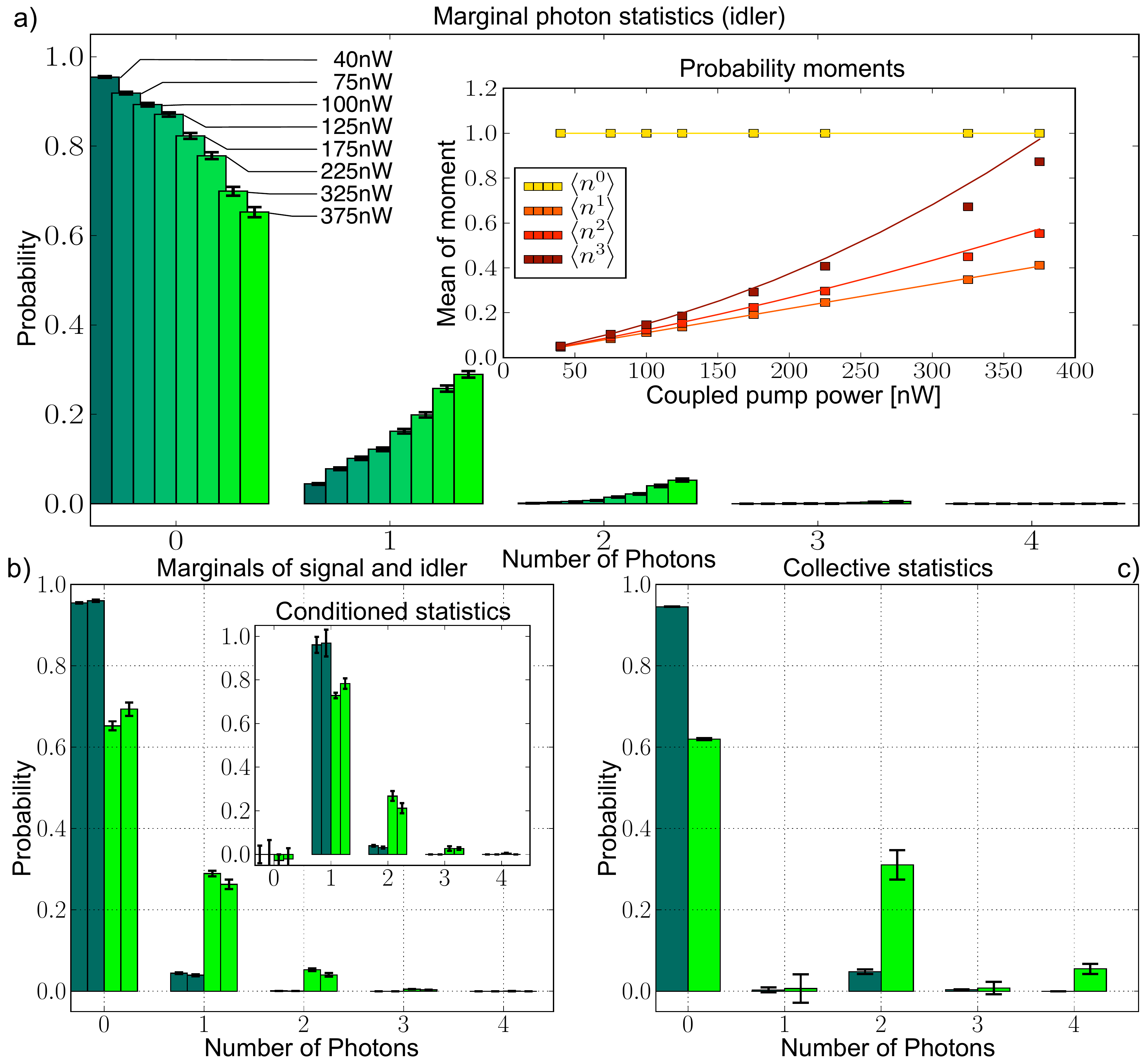}        
    \caption{(a) Marginal photon number statistics in idler mode for various 
                pump powers; Inset: first probability moments 
                $\langle  n^m \rangle$ with a fit to a Poissonian 
                distribution. (b) Comparison of signal and idler marginals 
                for lowest and highest pump power; Inset: conditioned statistics 
                demonstrate vacuum suppression. (c) Collective photon pair 
                statistics (see text) for lowest and highest pump powers.}
    \label{fig:marginals}
\end{figure}

In our second configuration, setup B, we study the marginal statistics of the
signal and idler mode. We replace, for example, the idler-APD from the
previous setup by a TMD in order to obtain the statistics of the idler mode. A
knowledge of $\eta_i$ is crucial for the loss tolerant reconstruction of
photon statistics. We infer $\eta_i$ from a coincidence measurement with the
remaining APD in the signal mode. We introduce an
additional fiber coupler to run this measurement and find a decline in
efficiency to $\eta_i = 11.3\%$. We extract all parameters of our model
directly from the measured data: the TMD click statistics $\vec{\varrho}$ on a
shot-by-shot basis; the efficiency $\eta$ from coincidences between idler-TMD
and signal-APD for loss tolerant reconstruction via $\mathbf{L}$; the
occupation probability $P_k$ of a particular TMD time-mode defining the
convolution $\mathbf{C}$. The reconstructed photon statistics are obtained
using a direct matrix inversion of Eq.\ref{eq:singleTMD} \cite{Achilles}. The
resulting marginals are shown in Figure \ref{fig:marginals}a for a range of
pump powers, displaying several photon statistics with respective values
$p_0,\ldots, p_4$, which correspond to the growth of mean photon numbers.
Figures of merit reflecting the same information are easily computed, such as
the probability moments $\langle n^m \rangle$, depicted in inset of Figure
\ref{fig:marginals}a and compared to a Poissonian fit. Reconstructed marginals
for TMD measurements in signal and idler beams are presented in Figure
\ref{fig:marginals}b. The marginals show a good agreement for all photon
numbers between signal and idler, which validates our photon number
reconstruction as this is expected for photon number correlation. The conditioned
statistics (Figure \ref{fig:marginals}b inset) are needed for the
calibration and confirm the preparation of heralded single photon states for
the lowest pump power as well as the good suppression of vacuum contributions
for all pump powers.

Using setup C, we extend our testing of the consistency of the inversion
method. In conjunction with the previous measurements, we now compare signal
marginals, idler marginals and the collective statistics of the twin beams.
This experiment comprises the collective coupling of both signal and idler
beams into a single TMD. Thus, the configuration renders a strict test for the
occurrence of photon pairs. The suppression of odd photon numbers, presented
in Figure \ref{fig:marginals}c, provides direct evidence for photon number
correlations. Clearly, this configuration cannot be used to measure
coincidences between signal and idler photons. We therefore used the
calibration of $\eta_s$ and $\eta_i$ from previous measurement configurations.
As expected, the photon statistics exhibit oscillations in the collective
probability distribution, arising from the pairwise generation of photons.
Further, comparing the collective statistics to the marginal distributions of
Figure \ref{fig:marginals}b we can prove that the distributions are consistent
across all measurements, because $ p^\text{collective}_{2n} =
p^\text{signal}_n = p^\text{idler}_n $ holds.

\begin{figure}[htbp]
     \centering
             \includegraphics[width=\linewidth]{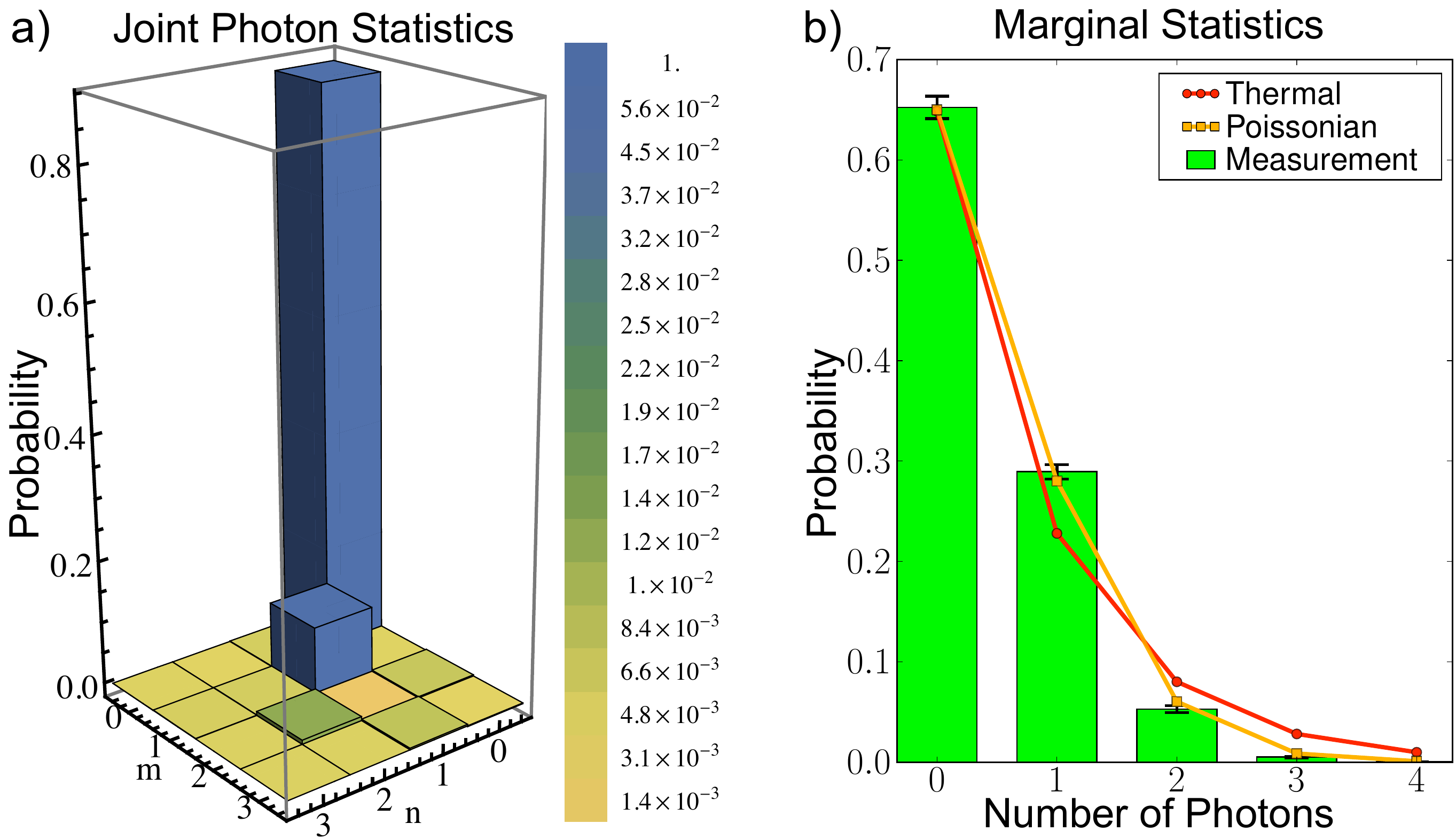}        
     \caption{(a) Joint photon statistics $p_{n,m}$ measured with two 
                TMDs and pump power of 220nW. (b) Marginal statistics of 
                idler with pump power of 375nW. The measurement is fitted 
                against thermal and Poissonian distributions.}
     \label{fig:joint_pnd}
 \end{figure}

We complement our state characterization with setup~D and demonstrate the
first direct joint photon number measurement by simultaneously operating two
TMDs, one for each beam. This enables us to measure directly all photon number
correlations between signal and idler. The theoretical model of a single TMD
can easily be extended to cover multiple TMDs. Let us assume that signal and
idler modes obey a photon number distribution for which $p_{n,m}$ is the
probability to find $n$ and $m$ photons in signal and idler beams
respectively. Each beam is analyzed by one TMD independently:
\begin{equation}
	\vec{\varrho}_{k,l} = \big( \mathbf{C}_s \mathbf{L}_s(\eta_s) \otimes \mathbf{C}_i \mathbf{L}_i(\eta_i) \big) \vec{p}_{n,m}
\end{equation}
Experimentally we find that the probability correlation matrix $p_{n,m}$ is
dominantly populated on the diagonal, which proves the strict photon number
correlation between signal and idler. This, in turn, is a strong justification
for exploiting the a priori assumption of pair generation for our source, used
for calibrating the detector efficiency. In this experiment we coupled the
signal beam into a single mode (SM)-fiber TMD, and the idler beam into multi
mode (MM)-fiber TMD. The electronics records correlated click statistics
between both TMDs. The coupling into SM-fibers comes along with a decrease of
detection efficiency to a level of $\eta_s = 2.74\%$ whereas $\eta_i = 11.1\%$
remains unchanged. Our setup is sufficiently loss tolerant, though, to cope
with the low efficiency. The result $p_{n,m}$ for a measurement with pump
power of 220nW is depicted in Figure \ref{fig:joint_pnd}. Slight uncertainties
in the measured click statistics due to finite acquisition time of 300s and
the modest efficiency in the signal mode contribute to minute errors found in
off-diagonal elements. From this data we determine a normalized covariance of
0.996 in the photon number statistics, compared to 0.052 in the raw data. The
reconstructed twin beam photon number squeezing $\frac{\Delta^2 (n_s -
n_i)}{\langle n_s \rangle \, \langle n_i \rangle}$ is found to be $-23.9$dB
for the photon number statistics and $-0.2$dB for the raw data.


At high power levels we find a mismatch between the observed photon statistics
and the predictions from a single-mode theory of the PDC process, equally
described as a two-mode squeezer. The evaluation of the quadratic Hamiltonian
gives for the signal and idler mode a thermal marginal distribution $p_n =
\frac{\langle n \rangle^n}{(1+\langle n \rangle)^{n+1}} $. In contrast, our
results obey a Poissonian marginal distribution $p_n = e^{-\langle n \rangle}
\frac{\langle n \rangle^n}{n!},$ as depicted in Figure \ref{fig:joint_pnd},
and illustrated by the fits. The thermal fit misses the observation in the
one-photon contribution by 0.0614, which is more than seven times larger than
the statistical error. This error is derived analytically by Gaussian error
propagation applied to Eq. \ref{eq:singleTMD} taking into account
uncertainties in each model parameter. The measurement uncertainties for the
efficiencies are 0.009, and for the click statistics a conventional square
root dependence is assumed. The discrepancy between the observed Poissonian
statistics and the thermal distribution of a two-mode squeezed state can only
be explained by a PDC process, introducing a multitude of different modes.
The convolution of many independent thermal statistics yields the observed
Poissonian statistics.

In summary, we have demonstrated the first explicit measurement of
correlations and complete joint photon number statistics of a PDC source. Our
approach to use a TMD for loss-tolerant state characterization enabled us to
study the marginal photon number distributions in detail. This revealed a
clear signature of a multimode structure of the source. We expect the
multimode structure of light to play an increasing role for the development of
advanced quantum key distribution protocols as it could be exploited for
implementing higher dimensional information coding \cite{Zhang:2008p3836}.
Multimode states need to be considered for developing hybrid quantum
communication schemes that include continuous variable states as well as
single photons \cite{Barreiro_Rohde}.

We thank Bruno Menegozzi for his support on electronics. This work was
supported by the EC under QAP funded by the IST directorate as Contract No
015848.




\end{document}